\documentclass[10pt,a4paper]{article}
\RequirePackage{amsmath,amssymb}
\RequirePackage[dvipsnames,usenames]{color}
\usepackage{cite}
\usepackage{fullpage}
\usepackage[british]{babel}
\usepackage[utf8]{inputenc}
\usepackage[T1]{fontenc}
\usepackage[final]{showkeys} 
\usepackage[bookmarks]{hyperref}
\usepackage{amsthm}
\usepackage{graphicx}
\usepackage{subfigure}
\usepackage{braket}
\usepackage{mathrsfs}
\usepackage{color}

\usepackage{bm}
\usepackage{amsfonts}
\usepackage{braket}

\newcommand{\bc}{\begin{center}}
\newcommand{\ec}{\end{center}}

\newcommand{\be}{\begin{eqnarray}}
\newcommand{\ee}{\end{eqnarray}}
\renewcommand{\d}{\mbox{${\rm d}$}}

\title{Turnaround physics beyond spherical symmetry}

\author{Andrea~Giusti\thanks{E-mail: agiusti@ubishops.ca}  
~and Valerio~Faraoni\thanks{E-mail: vfaraoni@ubishops.ca}
$\,$
\\
\\
{\em Department of Physics \& Astronomy, Bishop's University}
\\
{\em 2600 College Street, Sherbrooke Qu\'ebec, Canada J1M 1Z7}
}

\begin{document}
\def\theequation{\arabic{section}.\arabic{equation}} 

\maketitle
\begin{abstract}

The concept of turnaround surface in an accelerating universe is 
generalized to arbitrarily large deviations from spherical symmetry, to 
close the gap between the idealized theoretical literature and the real 
world observed by astronomers. As an analytical application, the 
characterization of turnaround surface is applied to small deviations from 
spherical symmetry, recovering a previous result while extending it to 
scalar-tensor gravity.

\end{abstract}

\newpage

\section{Introduction}
\label{sec:1}
\setcounter{equation}{0}

Thanks to the study of Type Ia supernovae \cite{Perlmutter:1997zf, 
Perlmutter:1998np}, it is known since 1998 that 
the present expansion of the universe is accelerated. In the context of 
general relativity, this acceleration is attributed to the 
presence of a 
mysterious form of dark energy permeating the universe and responsible for 
approximately seventy percent of its energy content  
\cite{AmendolaTsujikawabook}. However, since this dark 
energy is introduced completely {\em ad hoc}, there has been 
much activity in explaining the cosmic acceleration by modifying gravity 
at large scales and dispensing with dark energy 
(\cite{CCT, CDTT}, see Refs.~\cite{Sotiriou:2008rp, 
DeFelice:2010aj, Nojiri:2010wj, 
Capozziello:2011et, Capozziello:2009nq} for reviews). Whatever the 
explanation 
for the cosmic acceleration, there are physical phenomena peculiar to an 
accelerated 
universe. One of them is the {\em turnaround radius} of cosmic structures 
\cite{TR1,TR2,TR3,TR8}, 
which has attracted much attention recently \cite{TR4,TR5,TR7, 
TRA1, TRA2, TRA3, TRA4, TRA5, TRA6, TRA7, Lee:2015upn, Lee:2016oyu, 
Lee:2017ejv, Lee:2016qpt, KunzTR} 
because 
of its potential to test the $\Lambda$-Cold Dark Matter ($\Lambda$CDM) 
model and/or modified gravity. 

Consider an accelerating Friedmann-Lema\^itre-Robertson-Walker
(FLRW) universe and superpose a spherical matter condensation 
acting as a perturbation of the FLRW metric. The local gravitational 
attraction due to 
this overdensity tends to make it collapse, while the cosmic expansion 
tends to disperse it (if this structure is sufficiently large to feel the 
effect of the cosmic expansion appreciably). The turnaround radius is the 
minimum scale at which a spherical shell of test particles can not  
collapse 
because of the accelerated cosmic expansion (or, {\em vice-versa}, the 
upper limit to the radius of spherical bound structures in an accelerated 
universe). At 
the turnaround radius, the local attraction balances the cosmic expansion.

Thus far, with the exception of \cite{Turnandrea1, KunzTR}, the literature 
on turnaround physics has been restricted to spherically 
symmetric  situations. Ref.~\cite{Turnandrea1} studies analytically small 
deviations  from spherical symmetry (Ref.~\cite{KunzTR}, instead, examines 
larger deviations numerically). However, this idealized situation is still 
far from being realistic and can easily induce large errors (cf. 
Refs.~\cite{Lee:2015upn, Lee:2016oyu, Lee:2017ejv, Lee:2016qpt}).  
Currently,  the only reliable method to actually measure the turnaround point consists of  
using a pancake detection, as proposed in Ref.~\cite{Falco:2013bgy}, and then solving  
for zero velocity (see the discussion of Ref.~\cite{Lee:2015upn}).

In the presence of 
spherical symmetry, the 
turnaround radius trivially defines the ``turnaround surface'', {\em 
i.e.}, the sphere of radius equal to the turnaround radius, but 
the 
generalization of this turnaround surface to non-spherical  
situations of astrophysical interest has not been discussed in the 
theoretical literature. As a consequence, astronomers attempting to 
determine the turnaround surface and deduce cosmological  information 
have to grapple with ill-defined theory and basic concepts that are 
unclear, in addition to major observational 
challenges. 

Here we identify the salient features of the turnaround surface in 
spherical symmetry and characterize it with a definition suitable for 
geometries with arbitrarily large deviations from spherical symmetry 
(however, the deviations from the FLRW
metric remain always small). The 
key idea is to identify the turnaround surface with an equipotential 
surface of the (local) metric perturbation potential with the special 
property that, if test particles initially sit on this surface with zero 
velocity with respect to it, they remain on this surface at later times 
while it evolves. They must remain at rest on this surface and 
cannot leave it to collapse because of the self-gravity of the 
perturbation, nor disperse because of the cosmic expansion. 

There is only 
one critical surface ${\cal S^*}$ on which these two opposing forces 
balance, as in the spherically symmetric case.
Any other closed surface ${\cal S^*}$ nearby will 
not have this 
property: particles will collapse (if ${\cal S}^*$ is contained inside the 
critical surface) or will disperse, expanding faster than those at rest on 
the  critical 
turnaround surface, if ${\cal S}^*$ lies outside of it. This 
characterization captures the essence 
of turnaround 
sphere in spherical symmetry and generalizes this concept, while shifting 
the emphasis from the {\em size} of this surface (the turnaround radius) 
to the surface itself.

The study of the relevant equations for specific cosmic structures 
(observed or hypothetical) requires, in general, a numerical 
implementation. We can, however, apply our definition to an analytical 
discussion of small non-sphericities and test our characterization in this 
situation, which has already been studied in Ref.~\cite{Turnandrea1} with 
a completely different method, based on the splitting of the 
Hawking-Hayward quasilocal energy contained in the turnaround surface into 
local and cosmological parts. We recover the results of \cite{Turnandrea1} 
in our new, general description.

In Sec.~\ref{sec:2} we calculate the timelike geodesics needed in the 
rest of this paper, while Sec.~\ref{sec:3} provides the general definition 
of turnaround surface. The application to small non-sphericities is 
detailed in Sec.~\ref{sec:4}, while Sec.~\ref{sec:5} extends this 
result to scalar-tensor gravity. Sec.~\ref{sec:6} 
contains a discussion and the conclusions. We follow the notation of  
Ref.~\cite{Waldbook}.

\section{Timelike geodesics in the perturbed FLRW universe}
\label{sec:2}
\setcounter{equation}{0}

The definition of turnaround surface requires one to consider test 
particles 
lying on this surface. They follow  timelike geodesics in spacetime, 
therefore we first discuss these special curves traced by test 
particles and clouds (or shells) of dust in the perturbed FLRW spacetime.

The spacetime metric in the conformal Newtonian gauge is 
\be 
\d s^2 = g_{\alpha \beta} \, \d x^\alpha \d x^\beta =
a^2(\eta) \, \left\{ -(1 + 2 \Phi) \, \d \eta ^2 +
(1 - 2 \Phi) \Big[ \d r ^2 + r^2 \left( \d \theta ^2 + \sin ^2 
\theta \, \d \varphi ^2 \right) \Big]\right\} \,,
\label{perturbed-metric}
\ee
where $\eta$ is the conformal time of the unperturbed FLRW 
universe and $\Phi (x^i)$ describes the Newtonian perturbation. Since we 
will consider only structures of size much smaller than the Hubble radius 
$H_0^{-1}$, the time dependence of $\Phi$ can be safely neglected. 

Timelike geodesics parametrized by the proper time $\tau$ have 
four-tangents $u^{\mu}=dx^{\mu}/d\tau $ that satisfy the geodesic 
equation   
\be
\frac{du^{\mu}}{d\tau}+\Gamma^{\mu}_{\alpha\beta} \, u^{\alpha} u^{\beta} 
=0 \,. \label{minchia}
\ee
The only non-vanishing Christoffel symbols of the perturbed FLRW 
universe~(\ref{perturbed-metric}) are given in the appendix,  and they are 
used to compute the components of the timelike geodesic equation

\begin{equation} 
\begin{split}
\frac{\d u ^0}{\d \tau} = 
\frac{1}{a \, (2 \Phi + 1)} \Big\{ &a_{, \eta} \Big[2 \Big(r^2 (u^3)^2 
\sin^2 \theta +r^2 (u^2)^2 + (u^1)^2 - (u^0)^2\Big) \Phi \\ 
&-r^2 (u^3)^2 \sin ^2 \theta -r^2 (u^2)^2 - (u^1)^2 - (u^0)^2\Big]\\
&- 2 u^0 a \left(u^3 \Phi _{, \varphi} + u^2 \Phi _{, \theta}+ u^1 \Phi 
_{, r} \right) \Big\} \,,
\end{split} \label{geodesic-1}
\end{equation}

\begin{equation} 
\begin{split}
\frac{\d u ^1}{\d \tau} = 
\frac{1}{a (2  \Phi - 1)} \Big\{ &2 u^1 u^0 a_{, \eta} (1-2  \Phi) 
+a \Big[ \Big( r^2 (u^2)^2 + r^2 (u^3)^2 \sin ^2 \theta - (u^1)^2 + 
(u^0)^2 \Big)  \Phi _{, r}\\ 
&- 2 u^3 u^1  \Phi _{, \varphi} - 2 u^2 u^1 \Phi _{, \theta}+2 r \Big( 
(u^3)^2 \sin ^2 \theta + (u^2)^2 \Big) \Phi\\ 
&-r (u^3)^2 \sin ^2 \theta - r (u^2)^2 \Big] \Big\} \,, 
\end{split} \label{geodesic-2}
\end{equation}

\begin{equation} 
\begin{split}
\frac{\d u ^2}{\d \tau} = 
\frac{1}{r^2 a (2  \Phi - 1)} \Big\{ &2 r^2 u^2 u^0 a_{, \eta} (1-2 \Phi) 
+ a \Big[-2 r^2 u^2 u^3 \Phi _{, \varphi} \\
&+\Big(- r^2 (u^2)^2 + r^2 (u^3)^2 \sin ^2 \theta + (u^1)^2 + (u^0)^2 
\Big)  \Phi _{, \theta} \\
&- 2 r^2 u^1 u^2  \Phi _{, r} + r \Big(r (u^3)^2 \sin (2 \theta) - 4 u^1 
u^2\Big) \Phi \\ 
&- r^2 (u^3)^2 \sin \theta \cos \theta +2 r u^1 u^2\Big] \Big\} \,,
\end{split} \label{geodesic-3}
\end{equation}

\begin{equation} 
\begin{split}
\frac{\d u ^3}{\d \tau} = 
\frac{1}{r^2 a (2\Phi - 1)} \Big\{ &2 r^2 u^3 u^0 a_{, \eta} (1-2 \Phi) 
+a \Big[ \Big(\left(r^2 (u^2)^2+(u^1)^2+(u^0)^2\right) \csc ^2 \theta - 
r^2 (u^3)^2\Big)  \Phi _{, \varphi}\\ 
&+ 2 r u^3 \left(-r u^1  \Phi _{, r} -r u^2  \Phi _{, \theta} +r u^2 \cot 
\theta +u^1 \right) \\ 
&-4 r u^3 (r u^2 \cot \theta + u^1) \Phi \Big] \Big\} \,,
\end{split} \label{geodesic-4}
\end{equation}

In an unperturbed FLRW spacetime, the four-velocity 
of a comoving observer 
reads
\be
u^\mu _{(0)} = \left( u^0 _{(0)}, \bm{0} \right) = \left( \frac{1}{a}, 
\bm{0} \right) 
\ee
in coordinates $\left( \eta, r, \theta, \varphi \right)$. Adding a 
perturbation as in Eq.~(\ref{perturbed-metric}), the four-tangent to a 
timelike geodesic becomes
\be
 u^\mu = u^\mu _{(0)} + \delta u^\mu =  \left( u^0 _{(0)} + \delta u^0, 
\delta  \bm{u} \right) = 
\left( \frac{1}{a} + \delta u^0, \delta \bm{u} \right) \,; 
\label{stocazzo}
\ee
the normalization $u^\mu u_\mu = -1$ then yields 
\be
\delta u^0 = - \frac{\Phi}{a} 
\ee
to first order in $\delta u ^\mu$ and $\Phi$. Substituting this 
expression of $\delta u^0$ in the normalization of $u^\mu $, one finds
\be
-1 = - a^2 (1 + 2 \Phi) \left( \frac{1 - \Phi}{a} \right)^2 + g_{ij} \delta u^i \delta u^j = 
- 1 + g_{ij} \delta u^i \delta u^j + \mathcal{O} (\Phi ^2) \, .
\ee
If one then assumes that $\mathcal{O} (\delta u^1) = \mathcal{O} (\delta 
u^2) = \mathcal{O} (\delta u^3) $, then the latter 
implies
\be
\mathcal{O} (u^i) = \mathcal{O} (\delta u^i) = \mathcal{O} ( \Phi) \, , 
\quad i = 1, 2, 3 \, .
\ee
We can now use these results to study the geodesic equations to first 
order in the perturbation. In detail, it is easy to show that 
Eq.~(\ref{geodesic-1}) reduces to an identity to $\mathcal{O} 
(\Phi)$, 
whereas 
for the spatial components one finds
\be
\label{general-result-GR}
\frac{\d ( \delta u ^i)}{\d \tau} + \frac{2a_{, \eta}}{a^2} \, \delta 
u^i + g^{ij} \partial _j \Phi = 0 \, .
\ee

\section{General definition of turnaround surface}
\label{sec:3}
\setcounter{equation}{0}

Here we generalize the definition of turnaround surface to the case in 
which deviations from spherical symmetry can be arbitrarily large. An 
examination of the salient features of the turnaround sphere in spherical 
symmetry shows that this surface is an equipotential surface of the 
perturbation potential $\Phi$. This is a 
necessary property, but it is not sufficient to identify the turnaround 
surface.  We require the extra property that, if test 
particles  initially lay on this surface and have zero initial velocity 
{\em with respect 
to it}, they remain on this surface at later times as the latter evolves. 
These dust particles, and the surface with respect to which they are 
at rest, are not comoving with the background FLRW universe. They would 
be comoving only if the cosmic fluid of this FLRW background universe was 
dust, 
but this cannot be true because this universe accelerates. Therefore, 
these dust particles and the surface they lie on necessarily do not 
comove with the background. Furthermore, these particles are slowed down  
by the attraction of the local gravity due to the mass contained inside 
the turnaround surface, therefore they expand more slowly than the FLRW 
background.

This property is still not sufficient to identify the turnaround surface 
because many timelike geodesics cross the turnaround surface, but we 
further restrict to timelike geodesics that initially have  zero velocity 
with respect to this surface. Since they satisfy the timelike geodesic 
equation, which is of second order, assigning their initial position (on 
$\Sigma_{t_0}$) and initial velocity (vanishing with respect to $\Sigma_{t_0} 
$) 
specifies them completely. In other words, these massive test particles 
stay on $\Sigma_{t_0}$ initially and at all later times and expand more 
slowly than the accelerating FLRW background. Finally, to complete 
the identification of the initial surface $\Sigma_{t_0}$, we require that, 
on this surface, the attraction due to the Newtonian perturbation balances 
exactly the cosmic expansion, so that the acceleration of dust particles 
on this initial surface vanishes.

We require 
the {\em turnaround surface} $\Sigma_t$ at (comoving) time $t$ to be a  
two-dimensional, closed, simply connected surface that, at all times 
$t$, is an equipotential surface of the  perturbation $\Phi$ such that:

\begin{itemize}

\item[i)] The time evolution of the surface is such that the 
three-dimensional components of the tangent to the timelike geodesics 
crossing $\Sigma_t$ are {\em locally} proportional to the gradient 
$\bm{\nabla}\Phi$  (and therefore perpendicular to $\Sigma_t$ in the 
three-dimensional sense): 
\be
u^i |_{\Sigma_t} = \sigma (t) \, g^{ij} \partial _j \Phi 
|_{\Sigma_t} \, .\label{eq:3.1}
\ee 

\item[ii)] A dust particle initially comoving with the surface remains on 
this surface.\footnote{This particle is not comoving with the cosmic 
fluid because this surface is {\em not} comoving with the cosmic 
substratum.} In other words, if a dust particle is initially comoving 
{\em with 
$\Sigma_{t_0}$} (not with the FLRW background), namely 
\be
u^i  |_{\Sigma_{t_0}} =  \sigma (t_0) \, 
g^{ij} \partial _j \Phi | _{\Sigma_{t_0}} \,,
\ee  
then at a time $t>t_0$ its 
3-velocity will satisfy 
\be 
u^i |_{\Sigma_t} = \sigma (t) \, g^{ij} 
\partial_j \Phi | _{\Sigma_t} \,.
\ee 

\item[iii)] In an unperturbed FLRW universe, the (purely 
radial) acceleration of  a 
massive test particle is $\ddot{r}= \ddot{a} r/a$ (this rather 
intuitive result has been 
obtained many times in the literature, using various methods 
\cite{v1, v2, v3, v4, v5, v6, v7, v8, v9, v10, v11, v12, v13, v14, v15, 
v16, v17, v18, v19, v20, v21, v22, v23, v24, v25, v26, v27, v28, v29, 
v30, v31, v32, v33, v34, v35, PriceRomano, Leandros, 
FaraoniJacques}). 
In the presence of  a spherical perturbation, the turnaround radius is 
obtained by balancing the attraction of the local inhomogeneity with the 
cosmic 
acceleration, or $ \frac{M}{r^2}= \frac{ \ddot{a}}{a} \, r$, where $M$ is 
the mass of the local perturbation. For  a general Newtonian perturbation 
described by the potential $\Phi$, we impose that at every point of the 
initial surface $\Sigma_{t_0}$, assumed to be 
convex,\footnote{The surface is assumed to be convex to 
avoid pathological possibilities, such as a mass distribution with two 
or more centres far away from each other (which, technically, is  a mass 
distribution but has nothing to do with a mass concentration on the verge 
of collapsing under its own gravity).}
the acceleration of a massive test 
particle normal to this surface vanishes because the local attraction 
$-\bm{\nabla}\Phi$ balances exactly the force per unit mass due to the 
cosmic expansion {\em at that point} $\frac{ \ddot{a}(t_0)}{a(t_0)}\,  
\bm{x}_\perp $   (on $\Sigma_{t_0}$ there is no 
sideways acceleration due to the local gravity of the perturbation because 
$\Sigma_{t_0}$ is an equipotential surface of $\Phi$).  In other words, 
let $\bm{x}$  denote the position of a point on $\Sigma_{t_0}$ embedded in 
the $3$-dimensional spacelike
slice of the spacetime. Thus, denoting by
$$
\bm{n}  = \left. \frac{\bm{\nabla} \Phi}{|\bm{\nabla} \Phi|} 
\right|_{\Sigma_{t_0}} 
$$
the normal to $\Sigma_{t_0}$, the following condition must hold:
\be
-\bm{\nabla}\Phi =  \frac{ \ddot{a}(t_0)}{a(t_0)}\,  
\bm{x}_\perp \quad \mbox{on} \,\,\,  \Sigma_{t_0} \, , 
\ee
with $\bm{x}_\perp \equiv (\bm{x} \cdot \bm{n}) \, \bm{n}$, which implies
\be
-|\bm{\nabla}\Phi|^2 =  \frac{ \ddot{a}(t_0)}{a(t_0)}\,  
\bm{x} \cdot \bm{\nabla}\Phi 
\qquad \mbox{on} \;  \Sigma_{t_0} \,.
\ee
This condition completes the identification of the initial position of the 
massive test particles. 

\end{itemize}

To elucidate this definition let us consider some timelike dust 
comoving with the turnaround surface $\Sigma_{t_0}$ at some $t=t_0$. Since 
the dust is initially comoving with $\Sigma_{t_0}$, then the velocity of 
each particle will be such that $u^i |_ {\Sigma_{t_0} } \propto 
g^{ij} 
\partial _j \Phi | _{\Sigma_{t_0}}$, which implies $\delta u^i 
|_{\Sigma_{t_0} } =  \sigma(t_0) \, g^{ij} \partial _j \Phi 
|_{\Sigma_{t_0}}$. 
If this 
was not the case, a particle would have a non-vanishing component of $u^i | 
_{\Sigma_{t_0}}$ tangent to $\Sigma_{t_0}$, inducing a tangential 
movement along the surface, however we are not interested in this scenario 
with this definition. Furthermore, $\Phi$ is not constant in time: 
it does not depend explicitly on time, but it has a time-dependence through 
the coordinates on $\Sigma_t$, which depend on time,  $\Phi=\Phi 
\left( 
x^{\alpha}(t) \right)$.

Any other shell that does not satisfy precisely the two
initial conditions on position and velocity  
~i)~starting on $\Sigma_{t_0}$;
ii)~having zero initial velocity with respect to $\Sigma_{t_0}$ at $t_0$, 
and iii) coinciding with the zero acceleration surface at the 
initial time $t_0$, will
necessarily be forever distinct from $\Sigma_t$ due to the uniqueness 
theorem for the solutions of the Cauchy problem associated with the second 
order geodesic equation. The surface satisfying these two properties 
simultaneously is unique, and the definition of $\Sigma_t$ is  a true 
definition.

The turnaround surface deviates from the Hubble flow because of
the local attraction due to the mass contained in it, which creates the
first order metric perturbation potential. Consistently, in 
Eq.~(\ref{eq:3.1}), the induced
4-velocity of this shell {\em relative to the FLRW background} is of first
order. The turnaround surface $\Sigma_t$ evolves to background order
(as described by its 4-velocity $u^{\mu}_{(0)}$ {\em and} to first order, so
that its total 4-velocity is $u^{\mu}_{(0)}+\delta u^{\mu}$. The metric
perturbation potential $\Phi$ calculated on $\Sigma_t$ depends on the
coordinates on it, and this surface evolves, therefore also
$\Phi |_{\Sigma_t}$ {\em calculated on $\Sigma_t$} evolves, although it
has no explicit dependence on $t$ in the line 
element~(\ref{perturbed-metric}).

Now, the general result in \eqref{general-result-GR} can be recast as
\be
\label{general-result-GR-t}
\frac{\d ( \delta u ^i)}{\d t} + 2 H \delta 
u^i + g^{ij} \partial _j \Phi = 0 \, ,
\ee
since $u^0 _{(0)} = \d \eta / \d \tau = \d \eta / \d t = 1/a $ to 
order $\mathcal{O}(\Phi ^0)$. It is then easy to infer that 
\eqref{general-result-GR-t} reduces to
\be
\frac{1}{a^2} \frac{\d}{\d t}
\left( a^2 \, \delta u ^i \right)
= - g^{ij} \partial _j \Phi \, ,
\ee
leading to
\be
\delta u ^i | _{ \Sigma_t} = \frac{a^2 (t_0)}{a^2 (t)} \, \delta u ^i | 
_{\Sigma_{t_0}} - \frac{1}{a^2 (t)} \int _{t_0}^t h^{ij} (x^\alpha 
(t')) 
\partial _j \Phi (x^\alpha(t')) 
\, \d t' \,  
,\label{puralogica}
\ee
given that $ h^{ij} \equiv a^2(t) \, g^{ij} = \texttt{diag} (1, 
1/r^{2}, 
1/(r^{2}\sin^2 \theta))$. 

Astronomical observations of the turnaround radius cannot span the 
entire history of the structures observed since their formation, but only 
a small redshift interval  near the time when the light that is received now 
was  emitted by an object in the sky. Therefore, we linearize the quantities 
$a(t)$ and the integral to first order in $t-t_0$ (no astronomical 
observation has a chance to go beyond first order).

From the requirement of locality in (i), {\em i.e.} $\epsilon = t-t_0$ 
very small in comparison with $( H(t_0))^{-1}$, and expanding $a(t)= 
a(t_0)+H(t_0) \left(t-t_0\right) + \, ...$, one obtains
\begin{eqnarray}
\delta u ^i |_{ \Sigma_t} &=& 
\frac{1}{ \left[ 1+H(t_0) \left(t-t_0\right) 
+ \, ...\right]^2 } \, \delta u ^i |_{\Sigma_{t_0}} \nonumber\\
&&\nonumber\\
&\, &  -
\frac{1}{a^2(t_0) \left[ 1+H(t_0)\left(t-t_0\right) + \, ...\right]^2} \, 
\int_{t_0}^t \left[ h^{ij} \left( x^\alpha (t_0) +\dot{x}^{\alpha}(t_0) 
(t-t_0)\right)  \partial _j \Phi \left( x^\alpha(t_0) + 
\dot{x}^\alpha(t_0)(t-t_0) \right)\right]
\d t' \, \nonumber\\
&&\nonumber\\
&=&  \left[ 1 -2 H(t_0) \left(t-t_0\right)\right] 
\delta u ^i |_{\Sigma_0}  -
\frac{ \left[ 1 -2 H(t_0)\left(t-t_0\right) \right]}{ a^2(t_0)}   
 h^{ij} \left( x^\alpha (t_0)  \right) \partial_j \Phi \left( 
x^\alpha(t_0)\right) (t-t_0) + \, ... \,,
\end{eqnarray}
and finally
\be
\nonumber
\delta u ^i | _{\Sigma_t} &\!\!=\!\!& \left[\left( 1 - 2 H(t_0) \epsilon 
\right) \sigma(t_0)  - \frac{\epsilon}{a^2 (t_0)} \right] h^{ij} 
\partial 
_j \Phi | _{\Sigma_{t_0} } + \mathcal{O} (H^2(t_0) \epsilon^2) \\
&\!\!=\!\!& \left[\sigma(t_0) - \left( 2 \sigma(t_0) 
H(t_0) + 
\frac{1}{a^2 (t_0)}\right) \epsilon \right] h^{ij} \partial _j  \Phi 
|_{\Sigma_{ t_0}} + \mathcal{O} (H^2(t_0) \epsilon^2) \, .
\ee

	Consider now the 
spherical case. To first order, the equation of radial timelike geodesics 
reduces to (recall that $u^0 _{(0)} = \d \eta / \d \tau = 
\d \eta / \d t = 1/a $ to order $\mathcal{O}(\Phi ^0)$)
\be
\label{caso-sferico}
\frac{\d \delta u^1}{\d t} + 2 H \delta u^1 + \frac{\Phi'}{a^2} = 0 \, ,
\ee
where $\delta \bm{u}=\left( \delta u^1, 0, 0 \right) $, 
$\bm{\nabla}\Phi=  \left( \Phi', 0, 0 \right)$, $ \Phi= \Phi(r)$, and a 
prime denotes differentiation with 
respect to $r$. Clearly $\bm{u}$ and $\bm{\nabla}\Phi$ are parallel. 
As discussed above, Eq. \eqref{caso-sferico} can be recast as
\be
\frac{1}{a^2} \frac{\d}{\d t}
\left( a^2 \, \delta u ^i \right) = - \frac{\Phi'}{a^2} \, ,
\ee
which yields 
\be
\delta u^1 = \left[ \sigma(t_0) - \left( 2  \sigma(t_0) 
H(t_0) + 
\frac{1}{a^2 (t_0)}\right) (t - t_0) \right] \, \Phi' 
\equiv  \sigma (t) \, \Phi' (r) \, , 
\label{questa2}
\ee 
up to $\mathcal{O} (H^2(t_0) (t-t_0)^2)$.
Thus, we have proved that the only 3-velocity perturbation 
component $\delta u^1$ is 
proportional to the gradient of $\Phi$ on the turnaround sphere. 

In the coordinates $ \left( \eta, r, \theta, 
\varphi \right)$, this sphere evolves with time, hence, the 
proportionality constant $\sigma$ must depend on 
time (and only on time because this is an equipotential surface of the 
perturbation potential), as described by Eq.~(\ref{questa2}). Moreover, 
according to Eqs. \eqref{puralogica} and \eqref{questa2},
the velocity 
perturbation $\delta u^i$ is negative, which can be easily understood 
using  the spherical case as an example. The dynamics of the critical 
(turnaround) sphere was discussed in 
Ref.~\cite{usJCAP}: in an accelerated universe propelled by a cosmic fluid 
with equation of state $P=w\rho$, with $w\simeq -1 $ according to 
current observations, the areal radius $R_c$ of the critical turnaround 
sphere evolves according to \cite{usJCAP}
\be
\frac{ \dot{R}_c}{R_c} = \left( w +\frac{4}{3} \right) H \simeq 
\frac{H}{3} < H \,.
\ee  
This equation compares the expansion rate of the turnaround sphere 
with that of the cosmic substratum and tells us that the turnaround sphere 
expands slower than the Hubble flow. Therefore, a particle at rest on it 
will slow down with respect to the cosmic substratum and will have a 
radial velocity perturbation  $\delta u^1<0$.

\section{Small non-sphericities}
\label{sec:4}
\setcounter{equation}{0}

Let us apply now the previous considerations to small deviations from 
spherical symmetry. This situation is studied in \cite{Turnandrea1} 
with a conceptually different method. Ref.~\cite{Turnandrea1} is based on 
the splitting of the Hawking-Hayward quasilocal energy enclosed by a 
2-surface $\Sigma$ into a local and a cosmological part: the local part 
due to the perturbation $\Phi$ dominates inside the turnaround surface, 
while the cosmological part due to the cosmic mass-energy enclosed by 
$\Sigma $ dominates outside of the turnaround surface. This surface is 
defined by the equality of these two contributions (previously, this 
quasilocal energy method was 
applied to the spherical case \cite{usJCAP, usPRD}).  The result of 
Ref.~\cite{Turnandrea1} is that, to first order in the metric 
perturbations 
and in a parameter $\epsilon$ describing the deviations from spherical 
symmetry, the non-sphericities do not matter and the turnaround surface is 
still described by the turnaround radius obtained to zero order in 
$\epsilon$. Given the very different methods used in the present paper and 
in 
\cite{Turnandrea1}, one should check that the results obtained coincide 
and that the two methods are compatible.  Indeed, the results of 
\cite{Turnandrea1} are recovered in our new, general description of 
Sec.~\ref{sec:3}.

To wit: let us go back to the perturbed geodesic equations
\be
\frac{\d  (\delta u ^i)}{\d \tau} +  \frac{2a_{, \eta}}{a^2} \, \delta 
u^i + g^{ij} \partial _j \Phi = 0 \, .
\ee
Small non-sphericities are introduced by perturbing the otherwise 
spherical potential $\Phi_0(r)$ as 
\be
\Phi \left(r, \theta, \varphi \right) = \Phi _{(0)} \left(r \right) + 
\epsilon \, f \left( r, \theta, \varphi \right) \, , \quad\quad 
\mathcal{O} (\Phi) = \mathcal{O} (\Phi_0)  \, , \quad \,\,\, 0 
< \epsilon \ll 1 \, .  \label{NONS}
\ee
The non-sphericity  leads to a further correction in $\delta \bm{u}$, 
specifically 
\be
\delta \bm{u} = \left( \delta u^1 _{(0)}, 0, 0 \right) + \epsilon 
\, \bm{\delta} 
\ee
with $\delta  u^1 _{(0)}$ denoting the (radial) velocity perturbation in 
the unperturbed 
spherical case. Hence, perturbing the geodesic equations and using the 
results of the spherical case, one finds the equation satisfied by the 
non-sphericities 
\be
\frac{\d \delta^i}{\d t} + 2 H \delta^i + g^{ij} \partial _j f = 0
\ee
to $\mathcal{O}(\epsilon)$, where $H\equiv \dot{a}/a$ is the Hubble 
function with respect to comoving time. Finally, since the perturbed 
surface must be  
close to a sphere at all times, the perturbation function $f \left(r, 
\theta, \varphi \right)$ must be controlled so that it does not blow up. 
Unless some regulating condition is imposed to this regard, the 
function $f$ could grow very fast and the modified surface could deviate 
arbitrarily from a sphere even if the expansion parameter $\epsilon$ 
remains small. Therefore, we
impose that\footnote{This assumption has already been used in 
Ref.~\cite{Turnandrea1}.} $\bm{\nabla} f = 
\mathcal{O} (\epsilon)$,  then $\bm{\delta} = 
\mathcal{O} (\epsilon)$. This means that, for small deviations from 
sphericity, approximating the non-spherical turnaround surface with the 
unperturbed (spherical) one still gives the correct result to first order 
in the parameter $\epsilon$ that quantifies the non-sphericity in 
Eq.~(\ref{NONS}).

\section{Scalar-tensor gravity}
\label{sec:5}
\setcounter{equation}{0}

The turnaround radius has been studied also in scalar-tensor gravity and 
can, in principle, provide information about the theory of gravity at 
large 
scales \cite{TRA1, TRA2, TRA3, TRA4, TRA5, TRA6, KunzTR}. In scalar-tensor 
gravity there is a gravitational slip and a Newtonian  
perturbation describing a bound structure is described by two metric 
potentials $\Psi$ and $\Phi$. The perturbed FLRW line element is now
\be \label{STperturbed-metric}
\d s^2 = a ^2 (\eta) \left\{ -\left(1 + 2 \Psi \right) \, \d \eta ^2 +
 \left(1 - 2 \Phi \right) \Big[ \d r ^2 + r^2 \left(\d \theta ^2 + \sin ^2 
\theta \, \d \varphi ^2 \right) \Big] \right\}\,,
\ee
where it is assumed that the small metric perturbations $\Psi $ and $\Phi$ 
are time-independent and of the same order. The general definition of 
turnaround surface given above can still be applied, provided that this 
surface is now an equipotential surface of $\Psi$. 

Again, the only non-vanishing Christoffel symbols are given in the 
appendix and the equations of timelike geodesics with four-tangents 
$u^{\mu}$ are now

\begin{equation} 
\begin{split}
\frac{\d u ^0}{\d \tau} = 
\frac{1}{a \, (2 \Psi +1)} 
\Big\{ &a_{\eta} \Big[ 2 \Big(r^2 (u^3)^2 \sin ^2 \theta 
+ r^2 (u^2)^2+(u^1)^2\Big) \Phi - 2 (u^0)^2 \Psi \\
&- r^2 (u^3)^2 \sin ^2 \theta - r^2 (u^2)^2 -(u^1)^2 -(u^0)^2 \Big]\\
&-2 u^0 a \Big( u^3 \Psi_{, \varphi} + u^2 \Psi _{, \theta}  + u^1 \Psi_{, 
r} \Big)\Big\} \,,
\end{split}
\end{equation}

\begin{equation} \label{geodesic-2}
\begin{split}
\frac{\d u ^1}{\d \tau} = 
\frac{1}{a \, (2 \Phi -1)} 
\Big\{ &2 u^1 u^0 a _{,\eta} (1-2 \Phi) 
+ a \, \Big[ \Big( r^2 (u^2)^2 + r^2 (u^3)^2 \sin ^2 \theta - (u^1)^2 \Big) \Phi _{,r} \\
&-2 u^3 u^1 \Phi _{, \varphi} - 2 u^2 u^1 \Phi _{, \theta}
+2 r \Big( (u^3)^2 \sin ^2 \theta + (u^2)^2 \Big) \Phi \\ 
&+ (u^0)^2 \Psi _{,r} - r (u^3)^2 \sin ^2 \theta - r (u^2)^2\Big]\Big\} 
\,,
\end{split}
\end{equation}

\begin{equation} \label{geodesic-3}
\begin{split}
\frac{\d u ^2}{\d \tau} = 
\frac{1}{r^2 a (2 \Phi - 1)} \Big\{&2 r^2 u^2 u^0 a_{,\eta} (1 - 2 \Phi)
+a \, \Big[-2 r^2 u^2 u^3 \Phi _{, \varphi}\\
& + \Big( - r^2 (u^2)^2 + r^2 (u^3)^2 \sin ^2 \theta + (u^1)^2 \Big) \Phi _{, \theta}\\
&- 2 r^2 u^1 u^2 \Phi_{, r} 
+ r \Big( r (u^3)^2 \sin (2 \theta) - 4 u^1 u^2 \Big) \Phi \\ 
&+(u^0)^2 \Psi _{, \theta} - r^2 (u^3)^2 \sin \theta \cos \theta 
+2 r u^1 u^2
\Big]\Big\} \,,
\end{split}
\end{equation}

\begin{equation} \label{geodesic-4}
\begin{split}
\frac{\d u ^3}{\d \tau} = 
\frac{1}{r^2 a (2 \Phi -1)} \Big\{&2 r^2 u^3 u^0 a _{, \eta} (1 - 2 \Phi)
+a \Big[
\Big(\big(r^2 (u^2)^2+(u^1)^2\big) \csc ^2 \theta -r^2 (u^3)^2\Big) \Phi 
_{, \varphi}\\ 
& + 2 r u^3 \big( - r u^1 \Phi _{, r} - r u^2 \Phi _{, 
\theta} +r u^2 \cot \theta +  u^1\big)\\
&- 4 r u^3 (r u^2 \cot \theta + u^1) \Phi + (u^0)^2 \csc ^2 \theta \Psi _{, \varphi} 
\Big]\Big\} \,.
\end{split}
\end{equation}

The four-velocities $u^{\mu}=u^{\mu}_{(0)}+\delta u^{\mu}$ are given again 
by  Eq.~(\ref{stocazzo}). Specifically, to first order in both $\Psi$ and 
$\delta u ^\mu$, the 
normalization $u_{\mu}u^{\mu}=-1$ gives
\be \label{staminchia}
\delta u^0=-\frac{\Psi}{a} \,, \;\;\;\;\;\;\;\;
u^{\mu}=\left( \frac{1-\Psi}{a}, \delta u^1, \delta u^2, \delta u^3 
\right) \, .
\ee
Besides, plugging the latter again into the normalization of $u^\mu$ one finds
\be
-1 = - a^2 (1 + 2 \Psi) \left( \frac{1 - \Psi}{a} \right)^2 + g_{ij} \delta u^i \delta u^j = 
- 1 + g_{ij} \delta u^i \delta u^j + \mathcal{O} (\Psi ^2) \, ,
\ee
that implies $\delta \bm{u} = \mathcal{O} (\Psi) = \mathcal{O} (\Phi)$.

Substituting \eqref{staminchia} into the timelike geodesic equations and proceeding as done 
for general relativity in the previous sections, one can check that the 
time component of the geodesic equation is identically satisfied. The 
spatial components give, to first order:  
\begin{eqnarray}
&& \frac{d( \delta u^1)}{d\tau}+\frac{1}{a} \left( \frac{2a_{, \eta}}{a} 
\, \delta u^1 +\frac{\Psi_{,r}}{a} \right)=0 \,,\\
&&\nonumber\\
&& \frac{d( \delta u^2)}{d\tau}+\frac{1}{ar^2} \left( \frac{2r^2 a_{, 
\eta}}{a} \, \delta u^2 +\frac{\Psi_{,\theta}}{a} \right)=0 \,,\\
&&\nonumber\\
&& \frac{d( \delta u^3)}{d\tau}+\frac{1}{ar^2} \left( \frac{2r^2 a_{, 
\eta}}{a} \, 
\delta u^3 +\frac{\Psi_{,\varphi}}{a \, \sin^2\theta} \right)=0 \,.
\end{eqnarray}
Now we expand the potentials to describe small deviations from spherical 
symmetry as
\begin{eqnarray}
\Psi \left(r, \theta, \varphi \right)=
\Psi_0 \left(r \right) +\epsilon \, f \left(r, \theta, \varphi \right) 
\,,\\
&&\nonumber\\
\Phi \left(r, \theta, \varphi \right)=
\Phi_0 \left(r \right) +\epsilon \, h \left(r, \theta, \varphi \right) 
\,,
\end{eqnarray}
with $\epsilon$ a smallness parameter. Again, the four-velocities become
\be
u^{\mu}&\!\! = \!\!& u^{\mu}_{(0)} +\delta u^{\mu} = 
\left(  u^0_{(0)} +\epsilon  \, \delta_0 , 
\delta u^1_{(0)} +\epsilon \, \delta_1 , 
\epsilon \, \delta_2, \epsilon \, \delta_3 \right)\\
&\!\!=\!\!& \left( \frac{1-\Psi_0 -\epsilon\, f}{a} , \delta u^1_{(0)} 
+\epsilon \, 
\delta_1 , \epsilon \, \delta_2, \epsilon \, \delta_3 \right) \,.
\ee
Inserting this expansion into the spatial components of the geodesic 
equations yields
\begin{eqnarray}
&& \frac{d \delta_1}{dt}+ 2H\delta_1 =-\frac{f_{,r}}{a^2} \,,\\
&&\nonumber\\
&& \frac{d \delta_2}{dt}+ 2H \delta_1 
=-\frac{f_{,\theta}}{a^2r^2} \,,\\
&&\nonumber\\
&& \frac{d \delta_3}{dt}+ 2H \delta_3 
=-\frac{f_{,\varphi}}{a^2r^2\sin^2\theta} \,.
\end{eqnarray}
Again, one needs to control the behaviour of the deviations from 
sphericity by limiting the growth of the function $f$. This leads 
to the same results derived above for general relativity. One can conclude 
that, also in scalar-tensor gravity, 
small deviations from sphericity can be neglected in the identification of 
the turnaround surface.

\section{Discussion and conclusions}
\label{sec:6}
\setcounter{equation}{0}

In spherical symmetry, the turnaround radius clearly corresponds to a 
sphere of instability. Test particles that start on this 
surface with zero initial 
velocity with respect to it remain on it; analogous test particles inside 
this critical sphere must collapse, while those outside never form a 
bound system and disperse. The turnaround sphere corresponds to a 
delicate balance between the local gravitational attraction, which tends 
to make a dust shell collapse, 
and the cosmic expansion that tends to disperse it. On either side of the 
turnaround surface, one of these two forces prevails and moves a test 
particle away from it, so the position of (dynamical) equilibrium at the 
turnaround surface is clearly unstable. An actual measurement of the turn-around point will 
require the observation a specific galaxy near this surface. Therefore, this galaxy should preferably  
reside in a cold, coherent flow. It seems that this situation has only been studied in numerical 
simulations, with the conclusion that the needed galaxies are cold near the turnaround 
point \cite{Brinckmann:2014nia}.
The 4-velocity 
perturbation
$\delta u^1$ used in our calculation is negative because of the 
self-gravity of the mass 
contained inside the turnaround surface: this mass slows down the outward 
motion of geodesic particles relative to the cosmic substratum. In an 
unperturbed universe, massive test particles starting out with zero  
(radial) initial velocity relative to the background would not be pulled 
back this way,  hence it is always $\delta u^1 <0$ (assuming, of course, 
that the mass contained in the turnaround surface to be positive, which 
is the only physically meaningful option).

The turnaround radius is not  a fixed point in the phase space of radial 
timelike geodesics, unless the background universe is de Sitter, which is 
locally static \cite{PriceRomano, FaraoniJacques, Leandros}.  In a general 
FLRW 
background, the (proper or areal) turnaround radius is not constant  but 
depends on time and  the turnaround sphere expands (but the dust particles 
sitting on it have zero acceleration $\ddot{R}=0$, initially and at 
all later times, where $R$ is the areal 
radius). The turnaround sphere is not comoving.

Since the turnaround sphere is a sphere of unstable equilibrium, it marks 
the upper 
bound on the radius of any (spherical) bound structure. Because of the  
instability, a spherical bound structure with radius equal to the 
turnaround radius will not occur in nature. The turnaround radius is 
presented correctly in the literature as marking the upper limit to 
the largest possible size 
of a bound spherical structure.

Realistic structures in the sky, however, are not spherical nor  
approximately spherical. This fact is a challenge for  
astronomers attempting to identify the turnaround surface from 
observations of bound cosmic structures. This observational challenge is, 
of course, more complicated 
if one does not know what a turnaround surface is from the 
theoretical point of view. This is the gap addressed in the previous 
sections. 

In the absence of spherical symmetry, the ``size'' of an asymmetric bound 
structure, or cluster, may be defined operationally in various ways. Each 
one of them will have advantages and disadvantages and will be somehow 
questionable. However, it is more important to focus on the {\it 
turnaround surface} and to identify it, rather than discussing its 
``size''. Here we have identified the turnaround surface with 
an equipotential 
surface of the metric perturbation potential $\Phi$ satisfying a special 
property: dust particles initially sitting on this (non-spherical) 
surface with zero velocity with respect to it (and with the 
property that the gravitational acceleration due to the local mass 
distribution balances exactly the cosmic acceleration), will remain on it 
as this 
surface evolves in time. This definition is completely general and seems 
to be the correct generalization of 
turnaround sphere.  Then, the turnaround radius no longer 
exists\footnote{A ``turnaround 
size'' could be defined, for example, as $\ell_T = \sqrt{{\cal A}_T}$, 
where 
${\cal A}_T$ is the area of the turnaround surface defined above. In 
scalar-tensor gravity, this definition introduces the second potential 
$\Phi$ (in addition to $\Psi$) in $\ell_T$.} and the ``size'' 
of the critical turnaround surface ceases to play a primary role in the 
discussion of turnaround physics.

The application of our characterization of turnaround surface to small 
deviations from spherical symmetry reproduces the previous result of 
Ref.~\cite{Turnandrea1}, which was obtained with a completely different 
method (the splitting of the Hawking mass contained in the turnaround  
surface into local and cosmological  
parts \cite{usJCAP}). Further applications to realistic situations will be 
presented in the future.

\section*{Acknowledgments}
This work is supported by Bishop's University and by the Natural Sciences 
and Engineering Research Council of Canada (Grant No.~2016-03803 to V.F.).
The work of A.G. has been carried out in the framework of the activities of the Italian 
National Group for Mathematical Physics [Gruppo Nazionale per la Fisica Matematica (GNFM), 
Istituto Nazionale di Alta Matematica (INdAM)].

\section*{Appendix}

Here we report the only non-vanishing Christoffel symbols of the perturbed 
FLRW universe~(\ref{perturbed-metric}), which are used to compute the 
timelike geodesics in the text. They are

\begin{eqnarray}
 \Gamma ^0 _{00} &\!\!=\!\!& \frac{a _{, \eta}}{a} \, , \quad
\Gamma ^0 _{11} = \frac{a _{, \eta} (1 - 2 \Phi)}{a (1 + 2 \Phi )} \, , 
\quad
\Gamma ^0 _{22} = \frac{a_{, \eta} r^2 (1-2 \Phi)}{a (1 + 2 \Phi )} \, , \quad
 \Gamma ^0 _{33} = \frac{a_{, \eta} r^2 \sin ^2 \theta (1 - 2 \Phi)}{ a (1 
+ 2 \Phi)} \,, \\
 \Gamma ^0 _{01} &\!\!=\!\!& \frac{\Phi _{, r}}{1 + 2 \Phi} \, , \quad
 \Gamma ^0 _{02} = \frac{\Phi _{, \theta}}{1 + 2 \Phi} \, , \quad 
 \Gamma ^0 _{03} = \frac{\Phi _{, \varphi}}{1 + 2 \Phi} \,,\\
 \Gamma ^1 _{11} &\!\!=\!\!& \frac{\Phi _{, r} }{2 \Phi-1} \, , \quad
 \Gamma ^1 _{21} = \frac{\Phi _{, \theta}}{2 \Phi - 1} \, , \quad
 \Gamma ^1 _{22} = - \frac{r \left(r \Phi _{, r}+2 \Phi -1 \right)}{2 \Phi 
-1}\,, \\
 \Gamma ^1 _{31} &\!\!=\!\!& \frac{\Phi _{, \varphi}}{2 \Phi - 1} \, , 
\quad 
 \Gamma ^1 _{33} = -\frac{r \sin ^2 \theta \left(r \Phi _{, r} +2 
\Phi -1\right)}{2 \Phi - 1} \, , \quad
 \Gamma ^1 _{01} = \frac{a _{, \eta}}{a} \, , \quad
 \Gamma ^1 _{00} = \frac{\Phi _{, r}}{1-2 \Phi} \,,\\ 
 \Gamma ^2 _{11} &\!\!=\!\!& \frac{\Phi _{, \theta}}{ r^2 (1-2 \Phi)} \, , 
\quad
 \Gamma ^2 _{21} = \frac{-r \Phi _{, r} - 2 \Phi +1}{ r(1-2 \Phi)} \, , \quad
 \Gamma ^2 _{22} = \frac{\Phi _{, \theta}}{2 \Phi - 1} \, , \\
 \Gamma ^2 _{32} &\!\!=\!\!& \frac{\Phi _{, \varphi}}{2 \Phi-1} \, , \quad
 \Gamma ^2 _{33} = -\frac{\sin \theta  \left[ \sin \theta \Phi _{, \theta} 
+ \cos \theta (2 \Phi - 1) \right]}{2 \Phi -1} \, , \quad
 \Gamma ^2 _{02} = \frac{a _{, \eta}}{a} \,, \\
 \Gamma ^2 _{00} &\!\!=\!\!& \frac{\Phi _{, \theta}}{r^2 (1-2 \Phi)}\,, \\
 \Gamma ^3 _{11} &\!\!=\!\!& \frac{\csc ^2 \theta  \Phi _{, \varphi}}{r^2  
(1-2 \Phi)} \, , \quad
 \Gamma ^3 _{22} = \frac{\csc ^2 \theta  \Phi _{, \varphi}}{1-2 \Phi} \, 
, \quad \Gamma ^3 _{31} = \frac{-r \Phi _{, r}- 2 \Phi +1}{r(1-2 \Phi)}\,,\\
 \Gamma ^3 _{32} &\!\!=\!\!& \frac{\Phi _{, \theta} + \cot \theta  (2 
\Phi - 1)}{2 \Phi - 1} \, , \quad
 \Gamma ^3 _{33} = \frac{\Phi _{, \varphi}}{2 \Phi - 1} \, , \quad
 \Gamma ^3 _{03} = \frac{a _{, \eta}}{a} \,,\\ 
 \Gamma ^3 _{00} &\!\!=\!\!& \frac{\csc ^2  \theta \Phi _{, \varphi}}{r^2 
(1-2 \Phi)} \,.
\end{eqnarray}

When scalar-tensor gravity is considered, instead of GR, the line element 
is given by Eq.(\ref{STperturbed-metric}) instead 
of~(\ref{perturbed-metric}). In this case, the corresponding non-vanishing 
Christoffel symbols are

\begin{eqnarray}
 \Gamma ^0 _{00} &\!\!=\!\!& \frac{a _{, \eta}}{a} \, , \quad
\Gamma ^0 _{11} = \frac{a _{, \eta} (1 - 2 \Phi)}{a  (1 + 2 \Psi )} 
 \, , \quad \Gamma ^0 _{22} = \frac{a_{, \eta} r^2 (1-2 \Phi)}{a (1 + 
2 \Psi )}  \, , \quad
\Gamma ^0 _{33} = \frac{a_{, \eta} r^2 \sin ^2 \theta  (1 - 2 \Phi)}{a (1 
+ 2 \Psi)} \\
\Gamma ^0 _{01} &\!\!=\!\!& \frac{\Psi _{, r}}{1 + 2 \Psi} \, , \quad
\Gamma ^0 _{02} = \frac{\Psi _{, \theta}}{1 + 2 \Psi} \, , \quad 
\Gamma ^0 _{03} = \frac{\Psi _{, \varphi}}{1 + 2 \Psi} \\
 \Gamma ^1 _{11} &\!\!=\!\!& \frac{\Phi _{, r} }{2 \Phi-1} \, , \quad
 \Gamma ^1 _{21} = \frac{\Phi _{, \theta}}{2 \Phi - 1} \, , \quad
 \Gamma ^1 _{22} = - \frac{r \left(r \Phi _{, r}+2 \Phi -1\right)}{2 \Phi -1} \\
 \Gamma ^1 _{31} &\!\!=\!\!& \frac{\Phi _{, \varphi}}{2 \Phi - 1} \, , \quad
 \Gamma ^1 _{33} = -\frac{r \sin ^2 \theta \left(r \Phi _{, r} +2 \Phi -1\right)}{2 \Phi - 1} \, , \quad
 \Gamma ^1 _{01} = \frac{a _{, \eta}}{a} \, , \quad
\Gamma ^1 _{00} = \frac{\Psi _{, r}}{1-2 \Phi}  \\ 
 \Gamma ^2 _{11} &\!\!=\!\!& \frac{\Phi _{, \theta}}{r^2 (1-2 \Phi)} \, , \quad
 \Gamma ^2 _{21} = \frac{-r \Phi _{, r} - 2 \Phi +1}{ r(1-2 \Phi)} \, , \quad
 \Gamma ^2 _{22} = \frac{\Phi _{, \theta}}{2 \Phi - 1} \, , \\
 \Gamma ^2 _{32} &\!\!=\!\!& \frac{\Phi _{, \varphi}}{2 \Phi-1} \, , \quad
 \Gamma ^2 _{33} = -\frac{\sin \theta  \left[ \sin \theta \Phi _{, \theta}+ \cos \theta (2 \Phi - 1) \right]}{2 \Phi -1} \, , \quad
 \Gamma ^2 _{02} = \frac{a _{, \eta}}{a} \\
 \Gamma ^2 _{00} &\!\!=\!\!& \frac{\Psi _{, \theta}}{r^2 (1-2 \Phi)} \\
 \Gamma ^3 _{11} &\!\!=\!\!& \frac{\csc ^2 \theta  \Phi _{, \varphi}}{r^2 (1-2 \Phi)} \, , \quad
 \Gamma ^3 _{22} = \frac{\csc ^2 \theta  \Phi _{, \varphi}}{1-2 \Phi} \, , \quad
 \Gamma ^3 _{31} = \frac{-r \Phi _{, r}- 2 \Phi +1}{r(1-2 \Phi)} \\
 \Gamma ^3 _{32} &\!\!=\!\!& \frac{\Phi _{, \theta} + \cot \theta  (2 \Phi - 1)}{2 \Phi - 1} \, , \quad
 \Gamma ^3 _{33} = \frac{\Phi _{, \varphi}}{2 \Phi - 1} \, , \quad
 \Gamma ^3 _{03} = \frac{a _{, \eta}}{a} \\ 
\Gamma ^3 _{00} &\!\!=\!\!& \frac{\csc ^2 \theta \Psi _{, 
\varphi}}{r^2 (1-2 \Phi)} \,.
\end{eqnarray}


\clearpage
\begin{thebibliography}{99}

\bibitem{Perlmutter:1997zf} 
  S.~Perlmutter {\it et al.} [Supernova Cosmology Project Collaboration],
%  ``Discovery of a supernova explosion at half the age of the Universe and its cosmological 
%implications,''
  Nature {\bf 391}, 51 (1998)
  %doi:10.1038/34124
  [astro-ph/9712212].

\bibitem{Perlmutter:1998np} 
  S.~Perlmutter {\it et al.} [Supernova Cosmology Project Collaboration],
%  ``Measurements of Omega and Lambda from 42 high redshift supernovae,''
  Astrophys.\ J.\  {\bf 517}, 565 (1999)
  %doi:10.1086/307221
  [astro-ph/9812133].

\bibitem{AmendolaTsujikawabook}
 L.~Amendola and S.~Tsujikawa, {\em Dark Energy, Theory and Observations} 
(Cambridge University Press, Cambridge, 2010).

\bibitem{CCT} 
  S.~Capozziello, S.~Carloni and A.~Troisi,
%  ``Quintessence without scalar fields,''
  Recent Res.\ Dev.\ Astron.\ Astrophys.\  {\bf 1}, 625 (2003)
  [astro-ph/0303041].

\bibitem{CDTT} 
  S.M.~Carroll, V.~Duvvuri, M.~Trodden and M.S.~Turner,
%  ``Is cosmic speed - up due to new gravitational physics?,''
  Phys.\ Rev.\ D {\bf 70}, 043528 (2004)
  %doi:10.1103/PhysRevD.70.043528
  [astro-ph/0306438].

\bibitem{Sotiriou:2008rp} 
  T.P.~Sotiriou and V.~Faraoni,
 % ``$f(R)$ Theories Of Gravity,''
  Rev.\ Mod.\ Phys.\  {\bf 82}, 451 (2010)
  %doi:10.1103/RevModPhys.82.451
  [arXiv:0805.1726 [gr-qc]].

\bibitem{DeFelice:2010aj} 
  A.~De Felice and S.~Tsujikawa,
%  ``$f(R)$ theories,''
  Living Rev.\ Rel.\  {\bf 13}, 3 (2010)
  %doi:10.12942/lrr-2010-3
  [arXiv:1002.4928 [gr-qc]].

\bibitem{Nojiri:2010wj} 
  S.~Nojiri and S.D.~Odintsov,
%  ``Unified cosmic history in modified gravity: from F(R) theory to Lorentz non-invariant models,''
  Phys.\ Rept.\  {\bf 505}, 59 (2011)
  %doi:10.1016/j.physrep.2011.04.001
  [arXiv:1011.0544 [gr-qc]].
  
  \bibitem{Capozziello:2011et} 
  S.~Capozziello and M.~De Laurentis,
%  ``Extended Theories of Gravity,''
  Phys.\ Rept.\  {\bf 509}, 167 (2011)
  %doi:10.1016/j.physrep.2011.09.003
  [arXiv:1108.6266 [gr-qc]].

\bibitem{Capozziello:2009nq} 
  S.~Capozziello, M.~De Laurentis, and V.~Faraoni,
%  ``A Bird's eye view of f(R)-gravity,''
  Open Astron.\ J.\  {\bf 3}, 49 (2010)
  %doi:10.2174/1874381101003010049, 10.2174/1874381101003020049
  [arXiv:0909.4672 [gr-qc]].

\bibitem{TR1} 
  M.T.~Busha, F.C.~Adams, R.H.~Wechsler, and A.E.~Evrard,
%  ``Future evolution of structure in an accelerating universe,''
  Astrophys.\ J.\  {\bf 596}, 713 (2003)
  %doi:10.1086/378043
  [astro-ph/0305211].

\bibitem{TR2} 
  V.~Pavlidou and T.N.~Tomaras,
%  ``Where the world stands still: turnaround as a strong test of $\Lambda$CDM cosmology,''
  JCAP {\bf 1409}, 020 (2014)
  %doi:10.1088/1475-7516/2014/09/020
  [arXiv:1310.1920 [astro-ph.CO]].

\bibitem{TR3} 
  V.~Pavlidou, N.~Tetradis and T.N.~Tomaras,
%  ``Constraining Dark Energy through the Stability of Cosmic Structures,''
  JCAP {\bf 1405}, 017 (2014)
  %doi:10.1088/1475-7516/2014/05/017
  [arXiv:1401.3742 [astro-ph.CO]].

\bibitem{TR8} 
M.~Lapierre-L\'{e}onard, V.~Faraoni and F.~Hammad,
%  ``Cosmological applications of the Brown-York quasilocal mass,''
  Phys.\ Rev.\ D {\bf 96}, 083525 (2017)
  %doi:10.1103/PhysRevD.96.083525
  [arXiv:1710.06460 [gr-qc]].

\bibitem{TR4} 
S.~Bhattacharya and T.N.~Tomaras,
%  ``Cosmic structure sizes in generic dark energy models,''
  Eur.\ Phys.\ J.\ C {\bf 77}, 526 (2017)
  %doi:10.1140/epjc/s10052-017-5102-4
  [arXiv:1703.07649 [gr-qc]].

\bibitem{TR5} 
  M.~Cataneo and D.~Rapetti,
%  ``Tests of gravity with galaxy clusters,''
  Int.\ J.\ Mod.\ Phys.\ D {\bf 27},  1848006 (2018)
  %doi:10.1142/S0218271818480061
  [arXiv:1902.10124 [astro-ph.CO]].

\bibitem{TR7} 
  Z.~Roupas,
%  ``The Gravothermal Instability at all scales: from Turnaround Radius to Supernovae,''
  Universe {\bf 5}, 12 (2019)
  %doi:10.3390/universe5010012
  [arXiv:1809.07568 [gr-qc]].

\bibitem{TRA1}
  V.~Faraoni,
%  ``Turnaround radius in modified gravity,''
  Phys.\ Dark Univ.\  {\bf 11}, 11 (2016)
  %doi:10.1016/j.dark.2015.11.001
  [arXiv:1508.00475 [gr-qc]].

\bibitem{TRA2}
  S.~Bhattacharya, K.F.~Dialektopoulos, and T.N.~Tomaras,
%  ``Large scale structures and the cubic galileon model,''
  JCAP {\bf 1605}, 036 (2016)
 %doi:10.1088/1475-7516/2016/05/036
  [arXiv:1512.08856 [gr-qc]].

\bibitem{TRA3}
  S.~Bhattacharya, K.F.~Dialektopoulos, A.E.~Romano, C.~Skordis, and T.N.~Tomaras,
%  ``The maximum sizes of large scale structures in alternative theories of gravity,''
  JCAP {\bf 1707},  018 (2017)
  %doi:10.1088/1475-7516/2017/07/018
  [arXiv:1611.05055 [astro-ph.CO]].

\bibitem{TRA4} 
  S.~Nojiri, S.D.~Odintsov, and V.~Faraoni,
%  ``Effects of modified gravity on the turnaround radius in cosmology,''
  Phys.\ Rev.\ D {\bf 98}, 024005 (2018)
  %doi:10.1103/PhysRevD.98.024005
  [arXiv:1806.01966 [gr-qc]].
   
\bibitem{TRA5} 
  S.~Capozziello, K.F.~Dialektopoulos, and O.~Luongo,
%  ``Maximum turnaround radius in $f(R)$ gravity,''
  Int.\ J.\ Mod.\ Phys.\ D {\bf 28}, 1950058 (2018)
  %doi:10.1142/S0218271819500585
  [arXiv:1805.01233 [gr-qc]].

\bibitem{TRA6} 
  R.C.C.~Lopes, R.~Voivodic, L.R.~Abramo and L.~Sodr\'{e},
%  ``Turnaround radius in $f(R)$ model,''
  JCAP {\bf 1809}, 010 (2018)
  %doi:10.1088/1475-7516/2018/09/010
  [arXiv:1805.09918 [astro-ph.CO]].

\bibitem{TRA7} 
  R.C.C.~Lopes, R.~Voivodic, L.R.~Abramo and L.~Sodr\'e,
%  ``Relation between the Turnaround radius and virial mass in $f(R)$ model,''
  JCAP {\bf 1907}, 026 (2019)
  %doi:10.1088/1475-7516/2019/07/026
  [arXiv:1809.10321 [astro-ph.CO]].

\bibitem{Lee:2015upn} 
  J.~Lee, S.~Kim and S.C.~Rey,
%  ``A Bound Violation on the Galaxy Group Scale: The Turn-Around Radius of NGC 5353/4,''
  Astrophys.\ J.\  {\bf 815}, 43 (2015)
  %doi:10.1088/0004-637X/815/1/43
  [arXiv:1511.00056 [astro-ph.CO]].

  \bibitem{Lee:2016oyu} 
  J.~Lee and G.~Yepes,
%  ``Turning Around along the Cosmic Web,''
  Astrophys.\ J.\  {\bf 832}, 185 (2016)
  %doi:10.3847/0004-637X/832/2/185
  [arXiv:1608.01422 [astro-ph.CO]].

\bibitem{Lee:2017ejv} 
  J.~Lee,
%  ``Estimating the Turn-Around Radii of Six Isolated Galaxy Groups in the 
%Local Universe,''
  Astrophys.\ J.\  {\bf 856},  57 (2018)
  %doi:10.3847/1538-4357/aab358
  [arXiv:1709.06903 [astro-ph.CO]].

\bibitem{Lee:2016qpt} 
J.~Lee,
%  ``On the Universality of the Bound-Zone Peculiar Velocity Profile,''
  Astrophys.\ J.\  {\bf 832}, 123 (2016)
  %doi:10.3847/0004-637X/832/2/123
  [arXiv:1603.06672 [astro-ph.CO]].

\bibitem{KunzTR} 
  S.~H.~Hansen, F.~Hassani, L.~Lombriser and M.~Kunz,
%  ``Distinguishing cosmologies using the turn-around radius 
%near galaxy clusters,'' 
JCAP 01, 048 (2020) 
  [arXiv:1906.04748 [astro-ph.CO]].

\bibitem{Turnandrea1} 
  A.~Giusti and V.~Faraoni,
%  ``Turnaround size of non-spherical structures,''
  Phys.\ Dark Univ.\  {\bf 26}, 100353 (2019)
  %doi:10.1016/j.dark.2019.100353
  [arXiv:1905.04263 [gr-qc]].

\bibitem{Falco:2013bgy}
M.~Falco, S.H.~Hansen, R.~Wojtak, T.~Brinckmann, M.~Lindholmer, and S.~Pandolfi,
%``A new method to measure the mass of galaxy clusters,''
Mon. Not. Roy. Astron. Soc. \textbf{442}, 1887 (2014)
%doi:10.1093/mnras/stu971
[arXiv:1309.2950 [astro-ph.CO]].

\bibitem{Waldbook} R.M. Wald, {\em General Relativity} (Chicago University 
Press, Chicago, 1984).

\bibitem{v1} J. Pachner, {\em Phys. Rev.} {\bf 132}, 1837 (1963); {\em 
Phys. Rev. B} {\bf 137}, 1379 (1965).

\bibitem{v2} W.M. Irvine, {\em Ann. Phys.} {\bf 32}, 322 (1965).

\bibitem{v3} R.H. Dicke and P.J.E. Peebles, {\em Phys. Rev. Lett.} {\bf 
12}, 435 (1964).

\bibitem{v4} C. Callan, R.H. Dicke, and 
P.J.E. Peebles, {\em Am. J. Phys.} {\bf 33}, 105 (1965).

\bibitem{v5} P. D'Eath, {\em Phys. Rev. D} {\bf 11}, 1387 (1975).

\bibitem{v6} R.P.A. Newman and G.C. McVittie, {\em Gen. Rel. Grav.} {\bf 
14}, 591 (1982).

\bibitem{v7} R. Gautreau, {\em Phys. Rev. D} {\bf 29}, 198 (1984).

\bibitem{v8} P.A. Hogan, {\em Astrophys. J.} {\bf 360}, 315 (1990).

\bibitem{v9} B.C. Nolan, {\em J. Math. Phys.} {\bf 34}, 1 (1993).

\bibitem{v10} J.L. Anderson, {\em Phys. Rev. Lett.} {\bf 75}, 3602 (1995).

\bibitem{v11} W.B. Bonnor, {\em Mon. Not. R. Astr. Soc.} {\bf 282}, 1467 
(1996).

\bibitem{v12} A. Feinstein, J. Ibanez, and R. Lazkoz, {\em Astrophys. J.} 
{\bf 495}, 131 (1998).

\bibitem{v13} F.I. Cooperstock, V. Faraoni, and D.N. Vollick, 
{\em Astrophys. J.} {\bf 503}, 61 (1998).
 
\bibitem{v14} K.R. Nayak, M.A.H. MacCallum, and C.V. Vishveshwara, {\em 
Phys. Rev. D} {\bf 63}, 024020 (2000).

\bibitem{v15} V. Guruprasad, arXiv:gr-qc/0005090

\bibitem{v16} V. Guruprasad,  arXiv:gr-qc/0005014.

\bibitem{v17} G.A. Baker Jr., arXiv:astro-ph/0003152.

\bibitem{v18} A. Dominguez and J. Gaite, {\em Europhys. Lett.} {\bf 55}, 
458 (2001).

\bibitem{v19} T.M. Davis and C.H. Lineweaver, {\em AIPC} {\bf 555}, 348 
(2001).

\bibitem{v20} G.F.R. Ellis, {\em Int. J. Mod. Phys. A} {\bf 17}, 2667 
(2002).

\bibitem{v21} B. Bolen, L. Bombelli, and R. Puzio, {\em 
Class. Quant. Grav.} {\bf 18}, 1173 (2001).

\bibitem{v22} C. Stornaiolo, {\em Gen. Rel. Grav.} {\bf 34}, 2089 (2002).

\bibitem{v23} T.M. Davis, C.H. Lineweaver, and J.K. Webb, {\em Am. J. 
Phys.} {\bf 71}, 358 (2003).

\bibitem{v24} L. Lindegren and D. Dravins, {\em Astron. Astrophys.} {\bf 
401}, 1185 (2003).

\bibitem{v24} C.J. Gao, {\em Class. Quant. Grav.} {\bf 21}, 4805 (2004).

\bibitem{v25} T.M. Davis and C.H. Lineweaver, {\em Publ. Astr. Soc. Pac.} 
{\bf 21}, 97 (2004).

\bibitem{v26} D.P. Sheehan and V.G. Kriss, astro-ph/0411299.

\bibitem{v27} W.J. Clavering, {\em Am. J. Phys.} {\bf 74}, 745 (2006).

\bibitem{v28} Z.-H. Li and A. Wang, {\em Mod. Phys. Lett. A} {\bf 22}, 
1663 (2007).

\bibitem{v29} \O . Gr\o n and \O . Elgar\o y, {\em Am. J. Phys.} {\bf 75}, 
151 (2007).

\bibitem{v30} L.A. Barnes, M.J. Francis, J. B. James, and G.F. Lewis, 
{\em Mon. Not. R. Astr. Soc.} {\bf 373}, 382 (2006).

\bibitem{v31} R. Lieu and D.A. Gregory, arXiv:astro-ph/0605611.

\bibitem{v32} P.K.F. Kuhfittig, {\em Int. J. Pure Appl. Math.} {\bf 49},  
577 (2008).

\bibitem{v33} G.S. Adkins, J. McDonnell, and R.N. Fell, {\em Phys. Rev. D} 
{\bf 75}, 064011 (2007).

\bibitem{v34} D.L. Wiltshire, {\em New J. Phys.} {\bf 9}, 377 (2007).

\bibitem{v35} M. Sereno and P. Jetzer, {\em Phys. Rev. D} {\bf 75}, 064031 
(2007).

\bibitem{usJCAP} 
 V.~Faraoni, M.~Lapierre-L\'{e}onard and A.~Prain,
%  ``Turnaround radius in an accelerated universe with quasi-local mass,''
  JCAP {\bf 1510}, 013 (2015)
  %doi:10.1088/1475-7516/2015/10/013
  [arXiv:1508.01725 [gr-qc]].

\bibitem{usPRD}
  V.~Faraoni, M.~Lapierre-L\'{e}onard and A.~Prain,
%  ``Do Newtonian large-scale structure simulations fail to include relativistic effects?,''
  Phys.\ Rev.\ D {\bf 92}, 023511 (2015)
  %doi:10.1103/PhysRevD.92.023511
  [arXiv:1503.02326 [gr-qc]].

\bibitem{PriceRomano} 
  R.H.~Price and J.D.~Romano, 
%  ``In an expanding universe, what doesn't expand?,'' 
  Am. \ J. \ Phys. {\bf 80}, 376 (2012) 
  [arXiv:gr-qc/0508052].

\bibitem{FaraoniJacques} 
  V.~Faraoni and A.~Jacques,
%  ``Cosmological expansion and local physics,''
  Phys.\ Rev.\ D {\bf 76}, 063510 (2007)
  %doi:10.1103/PhysRevD.76.063510
  [arXiv:0707.1350 [gr-qc]].

\bibitem{Leandros} 
  S.~Nesseris and L.~Perivolaropoulos,
%  ``The Fate of bound systems in phantom and quintessence cosmologies,''
  Phys.\ Rev.\ D {\bf 70}, 123529 (2004)
  %doi:10.1103/PhysRevD.70.123529
  [astro-ph/0410309].

\bibitem{Brinckmann:2014nia}
T.~Brinckmann, M.~Lindholmer, S.H.~Hansen, and M.~Falco,
%``Zeldovich pancakes in observational data are cold,''
JCAP \textbf{04}, 007 (2016) 
%doi:10.1088/1475-7516/2016/04/007
[arXiv:1411.6650 [astro-ph.CO]].




\end{thebibliography}
\end{document}